\documentclass[english,11pt,a4paper]{article}
\usepackage[T1]{fontenc}
\usepackage[latin9]{inputenc}
\usepackage{amsmath}
\usepackage{amsthm}
\usepackage{amssymb}
\usepackage{graphicx}
\usepackage{wasysym}

\makeatletter
\numberwithin{equation}{section}

\usepackage{jheppub}
\usepackage{slashed}
\usepackage{physics}
\usepackage{simpler-wick}

\makeatother

\usepackage{babel}
\begin{document}
\title{ Shedding light on neutrino self-interactions with solar antineutrino
searches }

\abstract{Solar antineutrinos are absent in the standard solar model
prediction. Consequently, solar antineutrino searches emerge as a
powerful tool to probe new physics capable of converting neutrinos
into antineutrinos. In this study, we highlight that neutrino self-interactions,
recently gaining considerable attention due to their cosmological
and astrophysical implications, can lead to significant solar antineutrino
production. We systematically explore various types of four-fermion
effective operators and light scalar mediators for neutrino self-interactions.
By estimating the energy spectra and event rates of solar antineutrinos
at prospective neutrino detectors such as JUNO, Hyper-Kamiokande,
and THEIA, we reveal that solar antineutrino searches can impose stringent
constraints on neutrino self-interactions and probe the parameter
space favored by the Hubble tension. 

}

\author[a,b]{Quan-feng Wu} 
\author[a]{and Xun-Jie Xu}
\affiliation[a]{Institute of High Energy Physics, Chinese Academy of Sciences, Beijing 100049, China} 
\affiliation[b]{School of Physical Sciences, University of Chinese Academy of Sciences, Beijing 100049, China} 
\preprint{\today}  
\emailAdd{wuquanfeng@ihep.ac.cn} 
\emailAdd{xuxj@ihep.ac.cn}  
\maketitle

\section{Introduction}

The Sun, serving as the most intense natural source of neutrinos detected
on Earth, provides a unique opportunity to investigate neutrino properties
and probing underlying new physics~\cite{Maltoni:2015kca,Xu:2022wcq}.
The standard solar model, combined with our current understanding
of neutrinos, predicts that only neutrinos --- not antineutrinos
--- can be emitted by the Sun. Consequently, the detection of solar
antineutrinos would be compelling evidence of new physics, which has
motivated numerous experimental efforts in the search for solar antineutrinos~\cite{Super-Kamiokande:2002exp,SNO:2004eru,KamLAND:2011bnd,Borexino:2019wln,KamLAND:2021gvi,Super-Kamiokande:2020frs}.

Theoretically, solar antineutrinos can be produced through neutrino-antineutrino
oscillations~\cite{Pontecorvo:1957cp,Pontecorvo:1957qd}, which were
considered   as one of the potential solutions to the historical
``solar neutrino missing problem''~\cite{Davis:1968cp,Bahcall:1968hc}.
This  effect could be significant in the presence of large neutrino
magnetic moments and the solar magnetic field~\cite{Cisneros:1970nq,Okun:1986na,Lim:1987tk,Akhmedov:1988uk,Akhmedov2003,Akhmedov:2022txm}. 
 In addition, solar antineutrinos might also be produced via neutrino
decay or DM annihilation in the Sun --- see e.g.~\cite{Beacom:2002cb,Lehnert:2007fv,Rott:2012qb,Bernal:2012qh,Guo:2015hsy,Funcke:2019grs,Hostert:2020oui,Picoreti:2021yct}.

In this work, we propose that neutrino self-interactions, which are
challenging to probe in the laboratory~\cite{Blinov:2019gcj,Brdar:2020nbj,Deppisch:2020sqh,Berryman:2022hds}
but have recently attracted significant attention due to their cosmological
and astrophysical implications~\cite{Kreisch:2019yzn,Das:2022xsz,Bustamante:2020mep,Chang:2022aas,Venzor:2022hql,Venzor:2023aka,Berryman:2022hds,Esteban:2021tub,Fiorillo:2023cas,Fiorillo:2023ytr,Li:2023puz,RoyChoudhury:2020dmd,RoyChoudhury:2022rva,Ioka:2014kca,Blum:2014ewa,Shoemaker:2015qul,Carpio:2021jhu,Chen:2023vkq,Barenboim:2019tux},
might serve as another source of solar antineutrinos. 

Consider, for instance, a four-neutrino effective operator $\frac{1}{\Lambda^{2}}\nu^{\dagger}\nu^{\dagger}\nu\nu$,
where $\nu$ denotes the two-component Weyl spinor of a neutrino and
$\Lambda$ is the cut-off energy scale. With this operator in play,
a final-state neutrino in any process responsible for solar neutrino
production can be substituted with one antineutrino and two neutrinos
in the final state, implying the existence of an additional process
for solar antineutrino production.  More generally, for all possible
Lorentz-invariant four-neutrino operators (e.g., $\nu\nu\nu\nu$,
$\nu^{\dagger}\overline{\sigma}_{\mu}\nu\nu^{\dagger}\overline{\sigma}^{\mu}\nu$,
etc.), we can argue that at least one antineutrino can be produced
in similar processes.

Solar antineutrinos can be readily detected through inverse beta decay
(IBD) which benefits from well-developed technology for both event
identification and reconstruction. Given the advantages of IBD detection,
  we anticipate that next-generation neutrino detectors will exhibit
substantially improved  sensitivity to solar antineutrinos. 

Therefore, in this work we investigate the potential of probing neutrino
self-interactions through future solar antineutrino searches at 
JUNO~\cite{JUNO:2015zny}, Hyper-Kamiokande (HK)~\cite{Hyper-Kamiokande:2018ofw},
and THEIA~\cite{Theia:2019non}.   As our results will show, solar
antineutrino searches can provide the most competitive constraints
on neutrino self-interactions within a certain range of parameter
space. Notably,  the parameter space proposed by Ref.~\cite{Kreisch:2019yzn}
 for resolving the Hubble tension will be fully probed by solar antineutrino
searches. 

This article is structured as follows:  In Sec.~\ref{sec:Four-fermion},
we formulate all possible four-fermion effective operators for neutrinos
and relate them to solar antineutrino production. Then we calculate
the fluxes of solar antineutrinos produced through these operators,
take the oscillation effect into account, and determine the IBD event
rates at neutrino detectors. In Sec.~\ref{sec:Light-mediators},
we open up the effective vertices and investigate the impact of light
mediators involved in neutrino self-interactions. Finally we conclude
in Sec.~\ref{sec:Conclusion} and relegate some details to the appendix.

\section{Four-fermion effective operators\label{sec:Four-fermion}}

For neutrino self-interactions, we start from four-fermion effective
operators. The most general Lorentz-invariant four-fermion operators
that one can write down for Majorana\footnote{In this work, we only consider Majorana neutrinos because Dirac neutrinos
with strong self-interactions are severely constrained by cosmological
$N_{{\rm eff}}$~\cite{Luo:2020sho,Luo:2020fdt}.} neutrinos are 
\begin{align}
{\cal L}_{S} & =\frac{1}{\Lambda_{S}^{2}}(\nu\nu)(\nu\nu)+{\rm h.c.}\thinspace,\label{eq:-1}\\
{\cal L}_{S'} & =\frac{1}{\Lambda_{S'}^{2}}(\nu\nu)(\nu^{\dagger}\nu^{\dagger})\thinspace,\label{eq:-2}\\
{\cal L}_{V} & =\frac{1}{\Lambda_{V}^{2}}(\nu^{\dagger}\overline{\sigma}^{\mu}\nu)(\nu^{\dagger}\overline{\sigma}_{\mu}\nu)\thinspace,\label{eq:-3}\\
{\cal L}_{V'} & =\frac{1}{\Lambda_{V'}^{2}}(\nu^{\dagger}\overline{\sigma}^{\mu}\nu)(\nu\sigma_{\mu}\nu^{\dagger})\thinspace,\label{eq:-8}\\
{\cal L}_{T} & =\frac{1}{\Lambda_{T}^{2}}(\nu\sigma^{\mu\nu}\nu)(\nu\sigma_{\mu\nu}\nu)+{\rm h.c.}\label{eq:-4}
\end{align}
Here we formulate the interactions using two-component Weyl spinors
following the convention in  Ref.~\cite{Dreiner:2008tw}, including
for example $\nu\nu\equiv\nu_{a}\nu^{a}$, $\nu^{\dagger}\nu^{\dagger}\equiv\nu^{\dagger\dot{a}}\nu_{\dot{a}}^{\dagger}$,
  $\sigma^{\mu}=(1,\vec{\sigma})$, $\overline{\sigma}^{\mu}=(1,-\vec{\sigma})$,
$\sigma^{\mu\nu}\equiv\frac{i}{4}\left(\sigma^{\mu}\overline{\sigma}^{\nu}-\sigma^{\nu}\overline{\sigma}^{\mu}\right)$,
and $\overline{\sigma}^{\mu\nu}\equiv\frac{i}{4}\left(\overline{\sigma}^{\mu}\sigma^{\nu}-\overline{\sigma}^{\nu}\sigma^{\mu}\right)$.
We adopt Weyl spinors for the purpose of showing the distinction between
neutrinos and antineutrinos more explicitly. This approach can be
conveniently transformed to the more commonly used formalism  of
Dirac spinors, as detailed in Appendix~\ref{sec:Transform}.  The
four (anti-)neutrinos in Eqs.~\eqref{eq:-1}-\eqref{eq:-4} may possess
different flavors, which will be discussed in detail in Sec.~\ref{subsec:flavor}.

In addition to Eqs.~\eqref{eq:-1}-\eqref{eq:-4}, one might consider
the operator $(\nu\sigma^{\mu\nu}\nu)(\nu^{\dagger}\overline{\sigma}_{\mu\nu}\nu^{\dagger})$
but it vanishes according to the Fierz identities. Besides, one can
show that ${\cal L}_{S'}$ and ${\cal L}_{V'}$ can be transformed
to ${\cal L}_{V}$, and ${\cal L}_{T}$ can be transformed to ${\cal L}_{S}$.
These aspects are detailed in Appendix~\ref{sec:Transform}.

\begin{figure}
\centering

\includegraphics[width=0.8\textwidth]{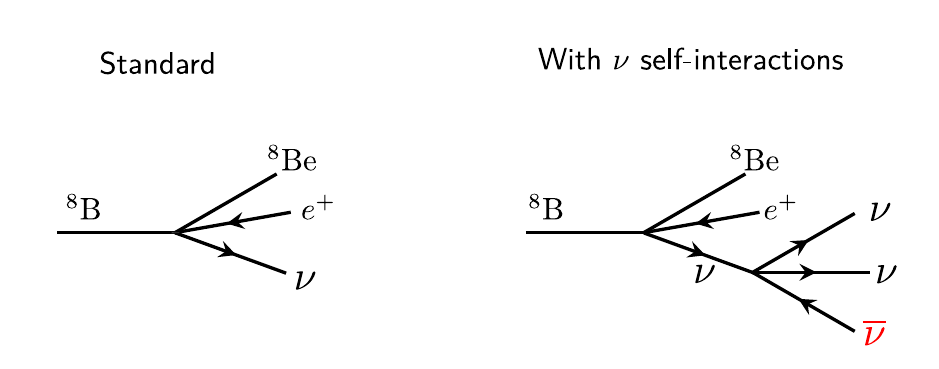}

\caption{Production of solar neutrinos (left panel) and antineutrinos (right
panel) from  $^{8}\text{B}$ decay in the presence of the $(\nu\nu)(\nu^{\dagger}\nu^{\dagger})$
operator. \label{fig:B8-decay}}

\end{figure}

\subsection{Solar antineutrino production \label{subsec:antineutrino-production}}

Let us now  consider the production of solar (anti)neutrinos. Solar
neutrinos are produced from fusion of light elements (e.g., $p+p\to\thinspace^{2}\text{H}+e^{+}+\nu_{e}$)
or decay of unstable elements created via fusion (e.g., $^{8}\text{B}\to{}^{8}\text{Be}+e^{+}+\nu_{e}$).
 In the presence of operators like $(\nu\nu)(\nu^{\dagger}\nu^{\dagger})$,
solar antineutrinos can be generated from processes\footnote{We note here that with the new processes involving neutrino self-interactions,
nuclear fusion and decay rates in the Sun are also slightly changed,
causing a small variation of energy production at the magnitude of
$\sim10^{-5}$ for the typical self-interaction strength considered
in this work. Given that the uncertainties in the density, temperature,
and chemical composition profiles are much higher than this magnitude,
such a variation is unlikely to be of  observational importance.} that split one neutrino from the aforementioned reactions into three
(anti)neutrinos, as depicted in Fig.~\ref{fig:B8-decay}. Obviously,
one of them must be an antineutrino if the splitting is caused by
the $(\nu\nu)(\nu^{\dagger}\nu^{\dagger})$ operator, which respects
the lepton number conservation. 

More generally, it is evident that other operators can also result
in the production of at least one antineutrino. Actually, the operators
$(\nu\nu)(\nu\nu)$ and $(\nu\sigma^{\mu\nu}\nu)(\nu\sigma_{\mu\nu}\nu)$
can produce three antineutrinos via analogous processes.  One might
speculate whether there could be operators consisting of one $\nu$
and three $\nu^{\dagger}$'s, which would imply that the right diagram
in Fig.~\ref{fig:B8-decay} produces no antineutrinos. However, due
to Lorentz invariance, this scenario is impossible. More specifically,
$\nu$ and $\nu^{\dagger}$ are under the $\left(\frac{1}{2},0\right)$
and $\left(0,\frac{1}{2}\right)$ representations of the Lorentz group
$\text{SO}(3,1)$, respectively, and there is no Lorentz-invariant
operator composed of an odd number of fields that are under $\left(\cdot,\frac{1}{2}\right)$
representations where ``$\cdot$'' standards for arbitrary integers
or half-integers. 

Hence, we conclude that any Lorentz-invariant four-neutrino effective
interactions inevitably lead to the production of  solar antineutrinos.

\subsection{Calculation of solar antineutrino spectra}

In this work, we only consider solar antineutrinos originating from
$^{8}\text{B}$ decay. In principle, pp fusion could, given the same
self-interaction strength,  produce a much higher antineutrino flux
than $^{8}\text{B}$ decay. However, the energy of an antineutrino
produced via pp fusion ($E_{\overline{\nu}}\lesssim0.4$ MeV) is well
below the threshold of IBD detection ($1.8$ MeV). So we do not consider
the contribution of pp fusion. 

The calculation of solar antineutrino spectra from $^{8}\text{B}$
decay involves the Feynman diagram in the right panel of Fig.~\ref{fig:B8-decay},
which requires the integration of five-body phase space. To our knowledge,
there is no simple analytical approach to this problem but some recursive
relations (see e.g. Ref.~\cite{byckling1973particle}) can be used
to facilitate numerical evaluations. We present the details in Appendix~\ref{sec:5-body}
and implement the algorithm in our code\footnote{Our code is publicly accessible via \url{https://github.com/Fenyutanchan/solar-vSI}.}.
The code allows us to perform  Monte Carlo integration of multi-body
phase space efficiently. 

\begin{figure}
\centering

\includegraphics[width=0.7\textwidth]{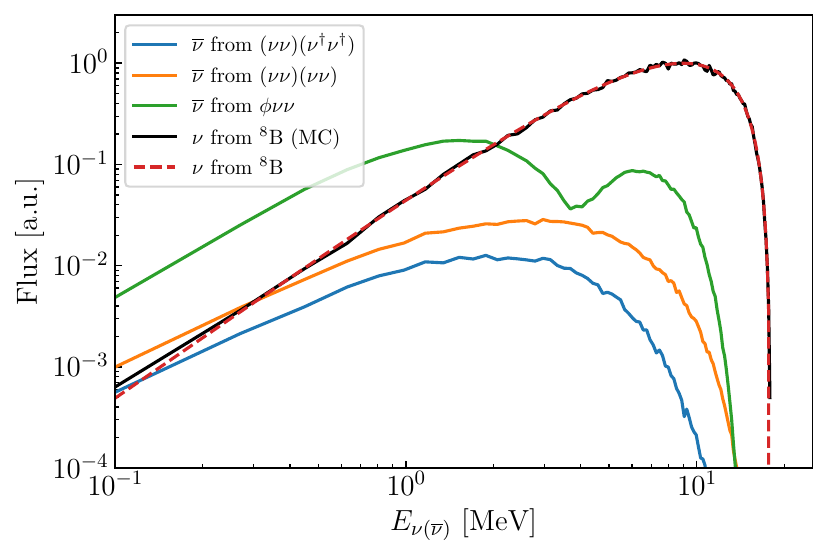}

\caption{Energy spectra of solar $\nu$ and $\overline{\nu}$ from $^{8}\text{B}$
decay.  All solid curves are obtained using the Monte Carlo method.
The dashed curve, which represents the standard $^{8}\text{B}$ decay,
is obtained from analytical calculation and has been normalized to
unity at the peak. The new physics  curves have been normalized by
the same factor multiplied by $10^{4}$.  The shown examples  take
$\Lambda_{S}=\Lambda_{S'}=10$ MeV and  $(m_{\phi}/{\rm MeV},\ g_{\phi})=(12.6,\ 1)$.
  \label{fig:Energy-spectra}}
\end{figure}

Figure~\ref{fig:Energy-spectra} shows the energy spectra of solar
$\nu$ and $\overline{\nu}$ from $^{8}\text{B}$ decay obtained using
our Monte Carlo code. Note that for the standard process $^{8}\mathrm{B}\to\mathrm{^{8}Be}+e^{+}+\nu_{e}$
which is a three-body decay, the energy spectrum can also be computed
analytically --- see Appendix~\ref{sec:3-body}. This is plotted
in Fig.~\ref{fig:Energy-spectra} as the red dashed curved. As is
shown in the figure, the Monte Carlo result of computing the standard
$^{8}\text{B}$ decay (black curve) agrees well with the analytical
result. 

In the presence of neutrino self-interactions, the antineutrino fluxes
from $^{8}\text{B}$ decay are shown by the blue, orange, and green
curves, for $(\nu\nu)(\nu^{\dagger}\nu^{\dagger})$, $(\nu\nu)(\nu\nu)$,
and $\phi\nu\nu$, respectively. The interaction $\phi\nu\nu$, together
with the double-peak structure of the flux, will be discussed in Sec.~\ref{sec:Light-mediators}.
The flux for $(\nu\nu)(\nu\nu)$ is higher than the flux for $(\nu\nu)(\nu^{\dagger}\nu^{\dagger})$.
This is because $(\nu\nu)(\nu\nu)$ converts one neutrino to three
antineutrinos while $(\nu\nu)(\nu^{\dagger}\nu^{\dagger})$ can only
produce one antineutrino. 

Here we would like to discuss an additional contribution to the antineutrino
flux. Given that neutrinos have three mass eigenstates and the lightest
one can be arbitrarily light, it is possible that a relatively heavy
neutrino state among them can decay into three lighter ones due to
the $4\nu$ self-interactions. Such decays could also produce antineutrinos.
However, due to the highly suppressed decay rates, this part of contribution
is negligible, as can be seen from the following estimate. For a mass
eigenstate with mass $m_{\nu}$ decaying to three massless states,
the decay rate can be estimated by analogy with the decay of muon\footnote{The muon decay rate is approximately given by $\Gamma_{\mu}\approx G_{F}^{2}m_{\mu}^{5}/192\pi^{3}$
if all final state masses are neglected.}:
\begin{equation}
\Gamma_{\nu\ {\rm decay}}\sim\frac{\Lambda^{-4}m_{\nu}^{5}}{192\pi^{3}}\sim\frac{1}{4\times10^{17}{\rm sec}}\left(\frac{{\rm MeV}}{\Lambda}\right)^{4}\left(\frac{m_{\nu}}{0.1\thinspace{\rm eV}}\right)^{5},\label{eq:nu-decay}
\end{equation}
which implies that the corresponding lifetime is almost as long as
the age of the universe ($4.35\times10^{17}$ sec) if $\Lambda=1$
MeV and $m_{\nu}=0.1$ eV. Taking this as a benchmark value, the probability
of a neutrino decaying before it arrives at Earth is 
\begin{equation}
P_{\nu\ {\rm decay}}\sim\Gamma_{\nu\ {\rm decay}}L_{\astrosun}\frac{m_{\nu}}{E_{\nu}}\sim10^{-23}\thinspace,\label{eq:P-decay}
\end{equation}
where $L_{\astrosun}=1.471\times10^{8}$ km is the distance from Earth
to the Sun, and $m_{\nu}/E_{\nu}\sim10^{-8}$ accounts for the relativistic
time dilation effect. Eq.~\eqref{eq:P-decay} implies that the antineutrino
flux produced via neutrino decays is  highly suppressed compared to
those presented in Fig.~\ref{fig:Energy-spectra}, hence negligible
in our analysis.

\subsection{Neutrino flavors and neutrino oscillations \label{subsec:flavor}}

\begin{figure}
\centering

\includegraphics[width=0.7\textwidth]{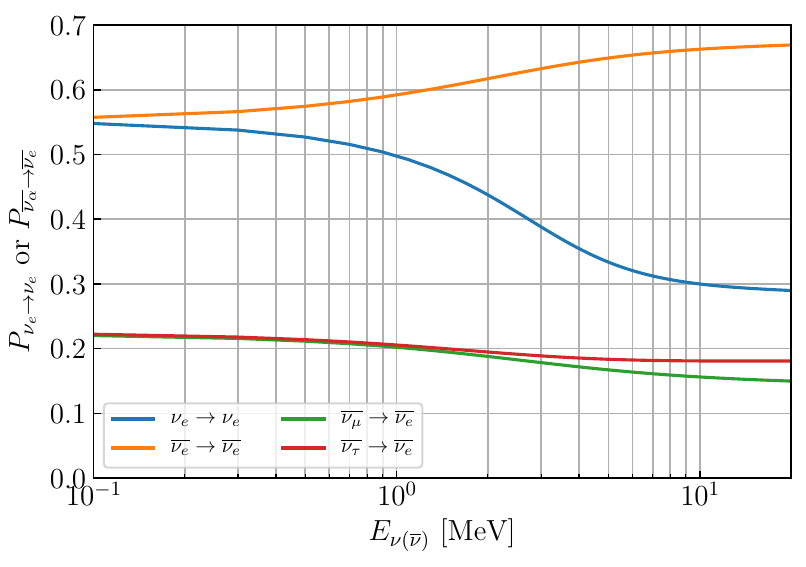}

\caption{Solar neutrino oscillation probabilities for $\nu_{e}\to\nu_{e}$
and $\overline{\nu_{\alpha}}\to\overline{\nu_{e}}$. \label{fig:osc}}
\end{figure}

We can generally assign different flavors to the (anti)neutrino fields
in Eqs.~\eqref{eq:-1}-\eqref{eq:-4}, e.g., $(\nu^{\dagger}\overline{\sigma}^{\mu}\nu)(\nu^{\dagger}\overline{\sigma}_{\mu}\nu)\to(\nu_{\alpha}^{\dagger}\overline{\sigma}^{\mu}\nu_{\alpha'})(\nu_{\beta}^{\dagger}\overline{\sigma}_{\mu}\nu_{\beta'})$.
Correspondingly, we add flavor indices to the $\Lambda$'s, e.g.,
~$\Lambda_{V}^{2}\to\Lambda_{V\alpha\alpha'\beta\beta'}^{2}$. Among
the various flavorful operators, only those with at least one $\nu_{e}$
are relevant to our analysis. So we only consider such operators and,
without loss of generality, we take $\alpha'=e$. 

The flavorful operators imply that antineutrinos being produced from
them may not necessarily be $\overline{\nu_{e}}$. They can be $\overline{\nu_{\mu}}$
or $\overline{\nu_{\tau}}$. Due to the standard neutrino oscillation,
they may be converted to $\overline{\nu_{e}}$ when arriving at the
Earth.  Note that the standard neutrino oscillation can change the
flavor composition, but not the composition of $\nu$ and $\overline{\nu}$. 

To account for the flavor conversion caused by neutrino oscillation,
we adopt the adiabatic approximation following the prescription in
Ref.~\cite{Xu:2022wcq} and calculate the conversion probabilities
of antineutrinos. The oscillation parameters used in our calculation
are~\cite{nu-fit,Esteban:2020cvm}:
\begin{align}
(\theta_{12},\ \theta_{13},\ \theta_{23},\ \delta_{{\rm CP}}) & =(33.41^{\circ},\ 8.54^{\circ},\ 49.1^{\circ},\ 0)\thinspace,\label{eq:-19}\\
(\Delta m_{21}^{2},\ \Delta m_{31}^{2}) & =(7.4\times10^{-5},\ 2.5\times10^{-3})\thinspace\text{eV}^{2}\thinspace.\label{eq:-20}
\end{align}
The result is illustrated in Fig.~\ref{fig:osc}. For $\nu_{e}\to\nu_{e}$,
the oscillation probability at the high-energy end of the curve is
approximately $\sin^{2}\theta_{12}\approx0.3$. But for $\overline{\nu_{e}}\to\overline{\nu_{e}}$,
the probability can be significantly higher, because antineutrinos
flip the sign of the MSW potential, resulting in $P_{\overline{\nu_{e}}\to\overline{\nu_{e}}}\approx\cos^{2}\theta_{12}\approx0.7$.
If the antineutrinos are initially produced as $\overline{\nu_{\mu}}$
or $\overline{\nu_{\tau}}$, neutrino oscillation can still convert
$\sim20\%$ of them to $\overline{\nu_{e}}$ according to Fig.~\ref{fig:osc}.
 Note that, unlike $\nu_{e}\to\nu_{e}$ or $\overline{\nu_{e}}\to\overline{\nu_{e}}$,
the flavor transition probabilities $P_{\overline{\nu_{\mu}}\to\overline{\nu_{e}}}$
and $P_{\overline{\nu_{\tau}}\to\overline{\nu_{e}}}$ depend on the
CP phase $\delta_{{\rm CP}}$ which is largely unknown. For neutrinos,
the effect of the CP phase on $P_{\nu_{e}\to\nu_{\alpha}}$ ($\alpha\neq e$)
is known to be significant, typically varying $P_{\nu_{e}\to\nu_{\alpha}}$
by $50\%\sim70\%$~\cite{Brdar:2023ttb}. In the case of antineutrinos,
our analysis reveals that this impact is considerably milder, confined
to roughly  $\sim10\%$. Hence, the influence of the CP phase can
be disregarded in our analysis.

\subsection{Event rates, $\chi^{2}$ analyses, and results}

By combing the Monte Carlo integration presented in Fig.~\ref{fig:Energy-spectra}
and the oscillation probabilities in Fig.~\ref{fig:osc}, it is straightforward
to compute the solar $\overline{\nu_{e}}$ flux arriving at the Earth.
Then we compute the event rates at three future neutrino detectors,
namely JUNO~\cite{JUNO:2015zny}, Hyper-Kamiokande (HK)~\cite{Hyper-Kamiokande:2018ofw},
and THEIA~\cite{Theia:2019non}. We consider the IBD detection channel
which is only sensitive to $\overline{\nu_{e}}.$  The event rates
are computed via

\begin{equation}
\frac{dN}{dE_{\nu}}=N_{\text{H}}\thinspace\Delta t\thinspace\sigma_{{\rm IBD}}\left(E_{\nu}\right)\phi_{\overline{\nu_{\alpha}}}(E_{\nu})P_{\overline{\nu_{\alpha}}\to\overline{\nu_{e}}}(E_{\nu})\thinspace,\label{eq:dN}
\end{equation}
where $N_{\text{H}}$ denotes the number of hydrogen atoms in the
detector, $\Delta t$ is the exposure time, $\sigma_{{\rm IBD}}$
is the cross section of IBD with a threshold of 1.8 MeV, and $\phi_{\overline{\nu_{\alpha}}}$
denotes the flux of $\overline{\nu_{\alpha}}$. For $\sigma_{{\rm IBD}}$,
we adopt the calculations in~\cite{Strumia:2003zx}. The exposure
time is universally assumed to be 10 years for all detectors. $N_{\text{H}}$
is computed from the fiducial mass of the detector, which for JUNO,
HK, and THEIA is 20, 187, and 100 kilo-tons, respectively. Since JUNO
is a liquid scintillator detector, we assume that the chemical composition
is $\text{C}_{{\rm n}}\text{H}_{2{\rm n}}$, in contrast to the other
two water-based ($\text{H}_{2}\text{O}$) detectors. 

\begin{figure}
\centering

\includegraphics[width=0.7\textwidth]{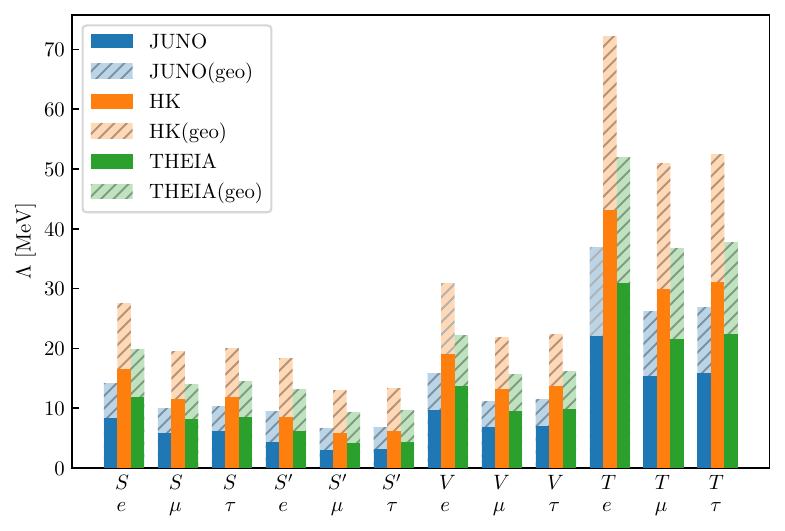}

\caption{The sensitivity reach of future experiments in probing the four-fermion
operators of neutrino self-interactions, at the $90\%$ C.L. The $x$-axis
labels ($S$, $S'$, $V$, $T$) indicate the type of the operators
--- see Eqs.~\eqref{eq:-1}-\eqref{eq:-4}, and $(e,\ \mu,\ \tau)$
indicate the flavor of the antineutrino being produced from $^{8}\text{B}$
decay. The bars in solid colors include reactor neutrinos and geoneutrinos
as backgrounds; the hatched bars in lighter colors assume the reactor-off
scenario, in which geoneutrinos become the dominant background. \label{fig:result-SVT}}
\end{figure}

To quantitatively evaluate the experimental sensitivity to neutrino
self-interactions, we adopt the following binned $\chi^{2}$ function~\cite{Zyla:2020zbs}:
\begin{equation}
\chi^{2}=\sum_{i}2\left(\mu_{i}-n_{i}+n_{i}\log\frac{n_{i}}{\mu_{i}}\right),\label{eq:-30}
\end{equation}
where $\mu_{i}$ and $n_{i}$ denote, respectively, the expected and
observed numbers of events in the $i$-th energy bin.  Eq.~\eqref{eq:-30}
allows us to readily include backgrounds, which should be added to
the signal as follows:
\begin{equation}
\mu_{i}=\mu_{i}^{{\rm \nu SI}}+\mu_{i}^{{\rm bkg}}\thinspace,\label{eq:-31}
\end{equation}
where $\mu_{i}^{{\rm \nu SI}}$ and $\mu_{i}^{{\rm bkg}}$ denote
the contributions of neutrino self-interactions and backgrounds, respectively. 

For solar antineutrino searches, the backgrounds mainly come from
reactor neutrinos and geoneutrinos. Other backgrounds such as the
diffuse supernova neutrino background (DSNB) and atmospheric neutrinos
within the energy range of interest are several orders of magnitude
lower so they are neglected in our analysis. For the reactor neutrino
background, we take the spectral shape from Ref.~\cite{Kopeikin:2012zz}
and the total fluxes, which are location-dependent, from Refs.~\cite{Xu:2022wcq,JUNO:2021vlw}.
For geoneutrinos, we assume a universal flux for all experiments,
taken from Ref.~\cite{Borexino:2019gps}.  Note that the reactor
neutrino background depends on the operational status of nearby reactors.
It is possible that the experiments may be able to collect a long
period of reactor-off data, either continuously or cumulatively, which
would feature  a substantially reduced reactor neutrino background.
To demonstrate the impact of this reduction, we also present results
without including the reactor background so that geoneutrinos become
the dominant backgroud. These are denoted by ``geo'' in the labels
in Fig.~\ref{fig:result-SVT}.  

When using Eq.~\eqref{eq:-30} to evaluate the experimental sensitivity,
 we set limits at the $90\%$ C.L., corresponding to $\Delta\chi^{2}=2.71$~\cite{Zyla:2020zbs},
where $\Delta\chi^{2}$ is the difference of the two $\chi^{2}$ values
with and without neutrino self-interactions. 

Our analyses for four-fermion effective operators are presented in
Fig.~\ref{fig:result-SVT}. The heights of the bars in Fig.~\ref{fig:result-SVT}
indicate the value of $\Lambda$  required to generate one IBD event
at the detector. In each instance, we consider only one type of the
operators, which can be $S$, $S'$, $V$, or $T$ according to Eqs.~\eqref{eq:-1}-\eqref{eq:-4}.
Results for $V'$ operators are not presented since they are identical
to the results for $V$. The flavor labels $(e,\ \mu,\ \tau)$ on
the $x$-axis indicate the initial flavor of the produced antineutrinos.
For simplicity, we assume that the antineutrinos originating from
$^{8}\text{B}$ decay are only of a single flavor. It is possible
that multi-flavor antineutrinos could be produced from one operator.
For instance, antineutrinos with the flavor composition $\overline{\nu_{e}}+\overline{\nu_{\mu}}+\overline{\nu_{\tau}}$
can be produced from the operator $(\nu_{\mu}\nu_{e})(\nu_{e}\nu_{\tau})$.
 In this case, one needs to sum over the flavor index  $\alpha$
in Eq.~\eqref{eq:dN}.

As is shown in Fig.~\ref{fig:result-SVT}, solar antineutrino observations
in future experiments might shed light on neutrino self-interactions
with $\Lambda$ at $\sim10$ MeV or even higher, depending on the
flavor and the type of interactions. This is to be compared with the
self-interaction strength favored by the Hubble tension in cosmology~\cite{Kreisch:2019yzn,Blinov:2019gcj}:
\begin{equation}
\Lambda_{H_{0}\ \text{tension}}=\begin{cases}
4.6\pm0.5\ \text{MeV} & (\text{SI})\\
90_{-60}^{+170}\ \text{MeV} & (\text{MI})
\end{cases}\thinspace,\label{eq:-21}
\end{equation}
where SI and MI indicates the \textquotedblleft strongly interacting\textquotedblright{}
(SI) or \textquotedblleft moderately interacting\textquotedblright{}
(MI) regimes considered in Ref.~\cite{Kreisch:2019yzn}. We see
that future solar antineutrino observations can fully probe the SI
regime and in some cases even enter the MI regime. 

\section{Light mediators\label{sec:Light-mediators}}

\subsection{Opening up the effective vertices}

As the order of magnitude of $\Lambda$ considered in the above analysis
is typically around  ten or a few tens of MeV, we should consider
how such operators might open up at energy scales comparable to $\Lambda$.
The simplest possibility would be adding a scalar ($\phi$) or vector
($A_{\mu}$) mediator such as\begin{equation}
(\nu\nu)(\nu\nu)
\to
 \wick{(\nu \c1 \phi \nu)(\nu \c1 \phi \nu)}
 ~~{\rm or } ~~
(\nu^{\dagger}\overline{\sigma}^{\mu}\nu)(\nu^{\dagger}\overline{\sigma}_{\mu}\nu)
\to
\wick{
(\nu^{\dagger}\overline{\sigma}\cdot \c2 A \nu)(\nu^{\dagger}\overline{\sigma} \cdot \c2 A \nu)
}\,.
\end{equation}  In addition, the four-fermion operators could also be generated
at the loop level via e.g. a box diagram. Here we investigate the
simplest scenario with a light real scalar mediator $\phi$ with the
following coupling and mass
\begin{equation}
{\cal L}\supset-\frac{1}{2}m_{\phi}^{2}\phi^{2}+\left(g_{\phi}\phi\nu\nu+{\rm h.c.}\right).\label{eq:-22}
\end{equation}
Here we assume that the coupling of $\phi$ to the electron is suppressed,
otherwise there would be much stronger laboratory limits.\footnote{If $\phi$ is coupled to the electron, it would be more efficiently
produced via thermal processes in the Sun and other stars, leading
to very restrictive stellar cooling bounds~\cite{Redondo:2008aa,An:2013yfc,Redondo:2013lna,Vinyoles:2015aba,An:2020bxd,Li:2023vpv}.
 } For such neutrinophilic light mediators, UV-complete models that
respect the SM gauge invariance can be built from e.g.~the right-handed
neutrino portal~\cite{Berbig:2020wve,Xu:2020qek,Chauhan:2020mgv,Chauhan:2022iuh}.

In the presence of the light scalar $\phi$, solar antineutrinos can
be directly produced from 

\begin{equation}
^{8}\text{B}\to{}^{8}\text{Be}+e^{+}+\overline{\nu}+\phi\thinspace,\label{eq:-24}
\end{equation}
or from the subsequent decay of $\phi$:
\begin{equation}
\phi\to\nu\nu\ (50\%)\ \text{or}\ \overline{\nu}\overline{\nu}\ (50\%)\thinspace,\label{eq:-25}
\end{equation}
where the percentages indicate the branching ratios of $\phi$ decay. 

The mean distance of $\phi$ traveling before decay is
\begin{equation}
\frac{\tau_{\phi}}{\sqrt{1-v^{2}}}=\frac{\tau_{\phi}E_{\phi}}{m_{\phi}}\sim0.1\ \text{mm}\cdot\left(\frac{10^{-3}}{g_{\phi}}\right)^{2}\cdot\left(\frac{{\rm MeV}}{m_{\phi}}\right)^{2}\cdot\frac{E_{\phi}}{10\ \text{MeV}}\thinspace,\label{eq:-23}
\end{equation}
where $\tau_{\phi}\sim16\pi/(g_{\phi}^{2}m_{\phi})$ is the lifetime
of $\phi$ at rest. Eq.~\eqref{eq:-23} implies that most $\phi$
particles being produced in the Sun decay within a sub-millimeter
distance. So we should include the contribution of Eq.~\eqref{eq:-25}
to the antineutrino flux. 

Interestingly, when the subsequent decay of $\phi$ is included, the
solar antineutrino spectrum may exhibit the double-peak structure
shown in Fig.~\ref{fig:Energy-spectra}, with the second peak arising
 from $\phi$ decay. We find that this feature is more significant
for heavier $\phi.$ This is because the energy of $\overline{\nu}$
produced from heavy $\phi$ decay  is generally higher than the mean
energy of $\overline{\nu}$ directly produced from $^{8}\text{B}$
decay. 

\subsection{Results}

\begin{figure}
\centering

\includegraphics[width=0.7\textwidth]{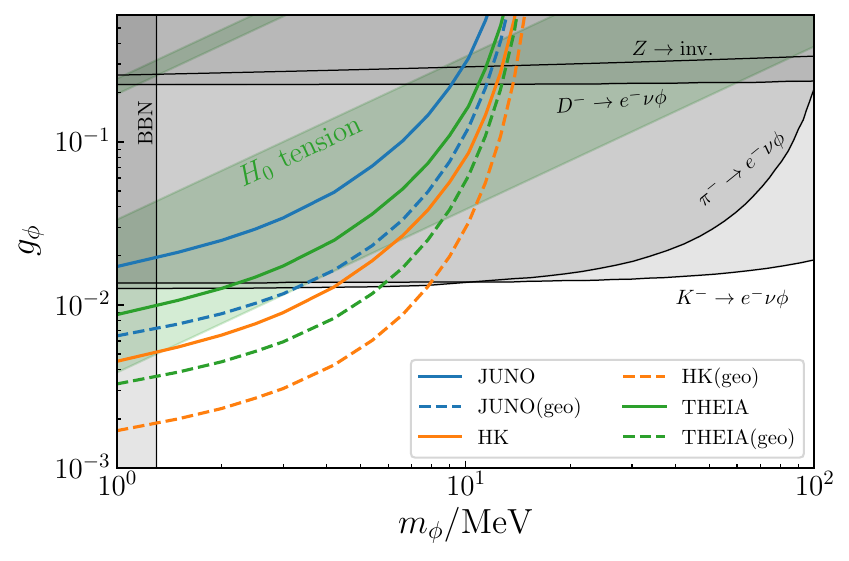}

\caption{The sensitivity reach of future experiments in probing the light scalar
$\phi$ as a mediator of neutrino self-interactions, at the $90\%$
C.L.  The solid lines include reactor neutrinos and geoneutrinos as
backgrounds; the dashed lines assume the reactor-off scenario, in
which geoneutrinos become the dominant background. \label{fig:result}}
\end{figure}

The antineutrino flux from $^{8}\text{B}$ decay combined with $\phi$
decay is calculated by employing the same Monte Carlo method utilized
in Sec.~\ref{sec:Four-fermion}. Following similar calculations,
we obtain the event rates at JUNO, HK, and THEIA.  We evaluate the
experimental limits on $g_{\phi}$ using the same $\chi^{2}$ function
in Eq.~\eqref{eq:-30} with $90\%$ C.L., and present them in Fig.~\ref{fig:result}.

Within the presented window, there are a few known bounds on neutrinophilic
scalars, derived from meson ($\pi^{-},$ $K^{-}$, $D^{-}$)  decay~\cite{Berryman:2018ogk},
$Z$ invisible decay~\cite{Brdar:2020nbj}, and Big Bang Nucleosynthesis
(BBN)~\cite{Blinov:2019gcj} --- all have been included in Fig.~\ref{fig:result}.
Compared to these known bounds, we see that solar antineutrino observations
can provide the most competitive constraints on neutrinophilic scalars
at the mass scale of a few MeV. 

In Fig.~\ref{fig:result}, we also present two green bands on the
plot, illustrating the self-interaction strengths favored by the Hubble
tension --- see Eq.~\eqref{eq:-21}. As is shown in the figure, when
combining all existing bounds on the light scalar $\phi$, the region
favored by the Hubble tension cannot be fully excluded, but the remaining
part of the space can be probed by solar antineutrino observations. 

Finally, we would like to mention  the possibility that the Hubble
tension may be resolved by other types of new physics or unidentified
systematic errors. In this situation, our work still demonstrates
that solar antineutrino observations can provide highly competitive
constraints on neutrino self-interactions  and also  serve as a valuable
complementary avenue for probing neutrino properties with cosmological
significance.

\section{Conclusion\label{sec:Conclusion}}

In this paper, we have explored the potential of solar antineutrino
observations as a probe of neutrino self-interactions, which could
have important implications for cosmology and astrophysics. We have
considered various scenarios of neutrino self-interactions including
all possible Lorentz-invariant four-fermion effective operators as
well as possible opening up of  the effective vertices by light mediators.
All these scenarios would lead to the production of solar antineutrinos.

Therefore, we present a dedicated calculation of the expected antineutrino
fluxes and event rates at neutrino detectors. We find that for the
self-interaction strength of cosmological interest, the resulting
antineutrino fluxes will cause significant solar IBD signals at next-generation
detectors like JUNO, HK, and THEIA.  In particular, we have demonstrated
that solar antineutrino spectra may exhibit distinctive features such
as double peaks due to the presence of a light mediator. 

Our results, as shown in Figs~\ref{fig:result-SVT} and \ref{fig:result},
imply that solar antineutrino observations can provide competitive
constraints on neutrino self-interactions, and in some cases fully
probe the parameter space favored by the Hubble tension. In conclusion,
solar antineutrinos offer a unique opportunity to test neutrino self-interactions
and might reveal unexplored features of neutrinos. 

\appendix

\section{Identities for Weyl, Dirac, and Majorana spinors\label{sec:Transform}}

This appendix compiles various known identities for Weyl, Dirac, and
Majorana spinors. We start by briefly reviewing a few identities for
anticommuting Weyl spinors $\chi_{i}$ ($i=1,$ 2, 3, $\cdots$)~\cite{Dreiner:2008tw}:
\begin{equation}
\chi_{1}\chi_{2}=\chi_{2}\chi_{1}\thinspace,\ \chi_{1}^{\dagger}\overline{\sigma}^{\mu}\chi_{2}=-\chi_{2}\sigma^{\mu}\chi_{1}^{\dagger}\thinspace,\ \chi_{1}\sigma^{\mu\nu}\chi_{2}=-\chi_{2}\sigma^{\mu\nu}\chi_{1}\thinspace,\ \chi_{1}^{\dagger}\overline{\sigma}^{\mu\nu}\chi_{2}^{\dagger}=-\chi_{2}^{\dagger}\overline{\sigma}^{\mu\nu}\chi_{1}^{\dagger}\thinspace,\label{eq:-7}
\end{equation}
where $\sigma^{\mu}=(1,\vec{\sigma})$, $\overline{\sigma}^{\mu}=(1,-\vec{\sigma})$,
$\sigma^{\mu\nu}\equiv\frac{i}{4}\left(\sigma^{\mu}\overline{\sigma}^{\nu}-\sigma^{\nu}\overline{\sigma}^{\mu}\right)$,
and $\overline{\sigma}^{\mu\nu}\equiv\frac{i}{4}\left(\overline{\sigma}^{\mu}\sigma^{\nu}-\overline{\sigma}^{\nu}\sigma^{\mu}\right)$.

In addition, there are several Fierz identities~\cite{Dreiner:2008tw}:
\begin{align}
(\chi_{1}\chi_{2})(\chi_{3}\chi_{4}) & =-(\chi_{1}\chi_{3})(\chi_{4}\chi_{2})-(\chi_{1}\chi_{4})(\chi_{2}\chi_{3})\thinspace,\label{eq:-9}\\
(\chi_{1}^{\dagger}\overline{\sigma}^{\mu}\chi_{2})(\chi_{3}^{\dagger}\overline{\sigma}_{\mu}\chi_{4}) & =2(\chi_{1}^{\dagger}\chi_{3}^{\dagger})(\chi_{4}\chi_{2})\thinspace,\label{eq:-10}\\
(\chi_{1}\sigma^{\mu\nu}\chi_{2})(\chi_{3}\sigma_{\mu\nu}\chi_{4}) & =-2(\chi_{1}\chi_{4})(\chi_{2}\chi_{3})-(\chi_{1}\chi_{2})(\chi_{3}\chi_{4})\thinspace,\label{eq:-11}\\
(\chi_{1}\sigma^{\mu\nu}\chi_{2})(\chi_{3}^{\dagger}\overline{\sigma}_{\mu\nu}\chi_{4}^{\dagger}) & =0\thinspace.\label{eq:-12}
\end{align}
Other Fierz identities not listed here are simply hermitian conjugates
of the above. According to the Fierz identities, it is straightforward
to see that ${\cal L}_{S'}$ and ${\cal L}_{T}$ can be transformed
to ${\cal L}_{V}$ and ${\cal L}_{S}$, respectively. In addition,
${\cal L}_{V'}$ can also be written into the form of ${\cal L}_{V}$
using Eq.~\eqref{eq:-7}. Therefore,only ${\cal L}_{S}$ and ${\cal L}_{V}$
are independent operators. 

To reformulate Eqs.~\eqref{eq:-1}-\eqref{eq:-4} in terms of Dirac
or Majorana spinors, we define
\begin{equation}
\psi_{D}\equiv\left(\begin{array}{c}
\nu\\
\eta^{\dagger}
\end{array}\right),\ \ \psi_{D}^{c}\equiv\left(\begin{array}{c}
\eta\\
\nu^{\dagger}
\end{array}\right),\ \ \psi_{M}\equiv\left(\begin{array}{c}
\nu\\
\nu^{\dagger}
\end{array}\right).\label{eq:-5}
\end{equation}
Here $\eta$ denotes the Weyl spinor of a right-handed neutrino. The
$4\times4$ Dirac matrices are connected to $2\times2$ Pauli matrices
as follows:
\begin{equation}
\gamma^{\mu}=\left(\begin{array}{cc}
 & \sigma^{\mu}\\
\overline{\sigma}^{\mu}
\end{array}\right),\ \ \sigma_{D}^{\mu\nu}\equiv\frac{i}{2}[\gamma^{\mu},\gamma^{\nu}]=2\left(\begin{array}{cc}
\sigma^{\mu\nu}\\
 & \overline{\sigma}^{\mu\nu}
\end{array}\right).\label{eq:-6}
\end{equation}

Using the above relations, we can reformulate Eqs.~\eqref{eq:-1}-\eqref{eq:-4}
in terms of Dirac spinors:
\begin{align}
{\cal L}_{S} & =\frac{1}{\Lambda_{S}^{2}}\overline{\psi_{D}^{c}}P_{L}\psi_{D}\overline{\psi_{D}^{c}}P_{L}\psi_{D}+{\rm h.c.}\thinspace,\label{eq:-1-1}\\
{\cal L}_{S'} & =\frac{1}{\Lambda_{S'}^{2}}\overline{\psi_{D}^{c}}P_{L}\psi_{D}\overline{\psi_{D}}P_{R}\psi_{D}^{c}\thinspace,\label{eq:-2-1}\\
{\cal L}_{V} & =\frac{1}{\Lambda_{V}^{2}}(\overline{\psi_{D}}\gamma^{\mu}P_{L}\psi_{D})(\overline{\psi_{D}}\gamma^{\mu}P_{L}\psi_{D})\thinspace,\label{eq:-3-1}\\
{\cal L}_{V'} & =\frac{1}{\Lambda_{V'}^{2}}(\overline{\psi_{D}}\gamma^{\mu}P_{L}\psi_{D})(\overline{\psi_{D}^{c}}\gamma^{\mu}P_{R}\psi_{D}^{c})\thinspace,\label{eq:-8-1}\\
{\cal L}_{T} & =\frac{1}{\Lambda_{T}^{2}}\frac{1}{4}(\overline{\psi_{D}^{c}}\sigma_{D}^{\mu\nu}P_{L}\psi_{D})(\overline{\psi_{D}^{c}}\sigma_{D\mu\nu}P_{L}\psi_{D})+{\rm h.c.}\label{eq:-4-1}
\end{align}
If using Majorona spinors, one simply needs to replace $\psi_{D}\to\psi_{M}$
and $\psi_{D}^{c}\to\psi_{M}$. Note that for a single neutrino flavor,
we have 
\begin{equation}
\overline{\psi_{M}}\gamma^{\mu}\psi_{M}=0\thinspace,\ \ \overline{\psi_{M}}\sigma_{D}^{\mu\nu}\psi_{M}=0\thinspace,\ \ \overline{\psi_{M}}\sigma_{D}^{\mu\nu}\gamma^{5}\psi_{M}=0\thinspace.\label{eq:-16}
\end{equation}
These can be obtained using Eq.~\eqref{eq:-7}.

\section{Calculation of three-body decay\label{sec:3-body}}

In this work, we encounter a three-body decay process: $^{8}\mathrm{B}\to\mathrm{^{8}Be}+e^{+}+\nu_{e}$.
To analytically calculate this process, we assume the following effective
Lagrangian:
\begin{equation}
{\cal L}\supset\frac{g_{{\rm eff}}^{2}}{M_{W}^{2}}(\Phi_{\mathrm{Be}}^{\dagger}\partial_{\mu}\Phi_{\mathrm{B}})(\overline{\psi_{\nu}}\gamma^{\mu}P_{L}\psi_{e})+\mathrm{h.c.},\label{eq:-26}
\end{equation}
where $g_{{\rm eff}}$ is a dimensionless coupling constant, $M_{W}$
is the mass of $W$ boson, $\Phi_{\mathrm{B}}$ and $\Phi_{\mathrm{Be}}$
describe $^{8}\mathrm{B}$ and $^{8}\mathrm{Be}$ as scalar fields,
and $\psi_{e}$ and $\psi_{\nu}$ denote the Dirac spinors of $e^{\pm}$
and $\nu_{e}$, respectively. Then the amplitude for $^{8}\mathrm{B}\to\mathrm{^{8}Be}+e^{+}+\nu_{e}$
reads
\begin{equation}
i{\cal M}(^{8}\mathrm{B}\to\mathrm{^{8}Be}+e^{+}+\nu_{e})=\frac{g_{{\rm eff}}^{2}}{M_{W}^{2}}\thinspace\overline{u_{3}}\slashed{p}P_{L}v_{2}\thinspace.
\end{equation}
where $p$ denotes the momentum of $^{8}$B; $u_{3}$ and $v_{2}$
denote the final states of $\nu_{e}$ and $e^{+}$ respectively. 

The amplitude squared, after applying the spin summation, reads
\begin{equation}
\begin{aligned}|{\cal M}(^{8}\mathrm{B}\to\mathrm{^{8}Be}+e^{+}+\nu_{e})|^{2} & =\frac{g_{{\rm eff}}^{2}}{M_{W}^{4}}{\rm tr}[\slashed{p}_{3}\slashed{p}P_{L}(\slashed{p}_{2}-m_{e})\slashed{q}P_{L}]\\
 & =\frac{g_{{\rm eff}}^{2}}{M_{W}^{4}}[4(p_{2}\cdot p)(p_{3}\cdot p)-2(p_{2}\cdot p_{3})p^{2}]\\
 & =\frac{2g_{{\rm eff}}^{2}M^{2}}{M_{W}^{4}}E_{2}E_{3}(1+\cos\theta_{23}),
\end{aligned}
\end{equation}
where $p_{1}$, $p_{2}$, and $p_{3}$ denote the four-momenta of
$^{8}$Be, $e^{+}$, and $\nu_{e}$, respectively. $M$ denotes the
mass of $^{8}$B.

Next, we shall compute the three-body phase integration, which reads
\begin{equation}
\begin{aligned}\int\dd{\Phi_{n}(p;p_{1},p_{2},p_{3})} & \equiv\int\frac{\dd[3]{\boldsymbol{p}_{1}}}{(2\pi)^{3}2E_{1}}\frac{\dd[3]{\boldsymbol{p}_{2}}}{(2\pi)^{3}2E_{2}}\frac{\dd[3]{\boldsymbol{p}_{3}}}{(2\pi)^{3}2E_{3}}(2\pi)^{4}\delta^{(4)}(p-p_{1}-p_{2}-p_{3})\\
 & =\frac{1}{8(2\pi)^{5}}\int\frac{\dd[3]{\boldsymbol{p}_{2}}\dd[3]{\boldsymbol{p}_{3}}}{E_{1}E_{2}E_{3}}\delta(E_{p}-E_{1}-E_{2}-E_{3})\\
 & =\frac{1}{8(2\pi)^{5}}\int\frac{E_{2}^{2}\dd{E_{2}}E_{3}^{2}\dd{E_{3}}}{E_{1}E_{2}E_{3}}\dd{\Omega_{2}}\dd{\Omega_{3}}\delta(E_{p}-E_{1}-E_{2}-E_{3})\\
 & =\frac{1}{4(2\pi)^{4}}\int\frac{E_{2}E_{3}}{E_{1}}\dd{E_{2}}\dd{E_{3}}\dd{\phi_{23}}\dd{\cos\theta_{23}}\delta(E_{p}-E_{1}-E_{2}-E_{3})\\
 & =\frac{1}{4(2\pi)^{3}}\int\dd{E_{2}}\dd{E_{3}}\dd{\cos\theta_{23}}\delta(E_{p}-E_{1}-E_{2}-E_{3})\frac{E_{2}E_{3}}{E_{1}}.
\end{aligned}
\end{equation}
Applying the above to $^{8}$B decay at rest, we obtain
\begin{equation}
\begin{aligned}\Gamma & =\frac{1}{2M}\int\dd{\Phi_{n}(q;k_{1},k_{2},k_{3})}\overline{\sum}\abs{\cal M}^{2}\\
 & =\frac{g_{{\rm eff}}^{2}M}{4(2\pi)^{3}M_{W}^{4}}\int\dd{E_{2}}\dd{E_{3}}\dd{\cos\theta_{23}}\delta(M-E_{1}-E_{2}-E_{3})\frac{E_{2}^{2}E_{3}^{2}}{E_{1}}(1+\cos\theta_{23})\\
 & =\frac{g_{{\rm eff}}^{2}M}{8(2\pi)^{3}M_{W}^{4}}\int\dd{E_{2}}\dd{E_{3}}\qty[M^{2}-M'^{2}-2M(E_{2}+E_{3})+4E_{2}E_{3}],
\end{aligned}
\end{equation}
where $M'$ is the mass of $^{8}\text{Be}$ and we have used 
\begin{align}
1+\cos\theta_{23} & =1+\frac{\qty(M-E_{2}-E_{3})^{2}-M'^{2}-E_{2}^{2}-E_{3}^{2}}{2E_{2}E_{3}}\nonumber \\
 & =\frac{M^{2}-M'^{2}-2M(E_{2}+E_{3})+4E_{2}E_{3}}{2E_{2}E_{3}}\thinspace,\\
\qty|\pdv{\qty(M-E_{1}-E_{2}-E_{3})}{\cos\theta_{23}}| & =\frac{E_{2}E_{3}}{E_{1}}\thinspace,
\end{align}
which are derived from 
\begin{equation}
M-E_{2}-E_{3}=E_{1}=\sqrt{M'^{2}+E_{2}^{2}+E_{3}^{2}+2E_{2}E_{3}\cos\theta_{23}}\thinspace.
\end{equation}

The differential decay rate is given by
\begin{equation}
\dv{\Gamma}{E_{3}}=\frac{g_{{\rm eff}}^{2}M}{8(2\pi)^{3}M_{W}^{4}}\int_{E_{2}^{\mathrm{min}}(E_{3})}^{E_{2}^{\mathrm{max}}(E_{3})}\dd{E_{2}}\qty[M^{2}-M'^{2}-2M(E_{2}+E_{3})+4E_{2}E_{3}],\label{eq:-17}
\end{equation}
where 
\begin{equation}
E_{2}(E_{3},\cos\theta)=\frac{M\qty(M-2E_{3})-M'^{2}}{2\qty[M-E_{3}\qty(1-\cos\theta_{23})]}\thinspace.
\end{equation}
It has the following upper and lower bounds:
\begin{align}
E_{2}^{\mathrm{min}}(E_{3}) & =\frac{M(M-2E_{3})-M'^{2}}{2M}\thinspace,\\
E_{2}^{\mathrm{max}}(E_{3}) & =\frac{M(M-2E_{3})-M'^{2}}{2(M-2E_{3})}\thinspace.
\end{align}
Integrating out $E_{2}$ in Eq.~\eqref{eq:-17}, we obtain
\begin{equation}
\begin{aligned}\dv{\Gamma}{E_{3}} & =\frac{g_{{\rm eff}}^{2}}{8(2\pi)^{3}M_{W}^{4}}\frac{E_{3}^{2}\qty[M(M-2E_{3})-M'^{2}]^{2}}{M(M-2E_{3})}\end{aligned}
\thinspace,
\end{equation}
where $E_{3}\in[0,\ (M^{2}-M'^{2})/(2M)]$. 

Further integrating  out $E_{3}$, we obtain the total decay rate:
\begin{equation}
\Gamma=\frac{g_{{\rm eff}}^{2}}{8(2\pi)^{3}M_{W}^{4}}\frac{1}{96M^{3}}\qty[M^{8}-8M^{6}M'^{2}+24M^{4}M'^{4}\ln\frac{M}{M'}+8M^{2}M'^{6}-M'^{8}].
\end{equation}

\section{Calculation of four- and five-body decay\label{sec:5-body}}

For $^{8}\mathrm{B}\to\mathrm{^{8}Be}+e^{+}+\overline{\nu}+\phi$
and $^{8}\mathrm{B}\to\mathrm{^{8}Be}+e^{+}+3\nu(\overline{\nu})$,
we need to compute four- and five-body phase space integrals. The
practical way of handling with such phase space integrals is to adopt
the time-like recursion relations~\cite{byckling1973particle}, which
will be briefly reviewed below. 

In general, an $n$-body decay rate reads
\begin{equation}
\dd{\Gamma}=\frac{1}{S}\frac{1}{2M}\abs{\cal M}^{2}\dd{\Phi_{n}},\label{eq:differential-decay-rate}
\end{equation}
where $S$ is the \emph{symmetry factor} accounting for identical
particles, $M$ is the mass of the initial particle, ${\cal M}$
is the amplitude of the decay process, and $\dd{\Phi_{n}}$ is given
by
\begin{equation}
\dd{\Phi_{n}}=(2\pi)^{4}\delta^{(4)}\qty(p-\sum_{i=1}^{n}p_{i})\prod_{i=1}^{n}\frac{\dd[3]{\boldsymbol{p}_{i}}}{(2\pi)^{3}2E_{i}}\thinspace.\label{eq:n-body-phase-space-element}
\end{equation}

The phase space can be written recursively as 
\begin{equation}
\begin{aligned}\dd{\Phi_{n}} & =\frac{\dd[3]{\boldsymbol{p}_{n}}}{(2\pi)^{3}2E_{n}}\qty(\prod_{i=1}^{n-1}\frac{\dd[3]{\boldsymbol{p}_{i}}}{(2\pi)^{3}2E_{i}})(2\pi)^{(4)}\qty(\qty(p-p_{n})-\sum_{i=1}^{n-1}p_{i})\\
 & =\frac{\dd[3]{\boldsymbol{p}_{n}}}{(2\pi)^{3}2E_{n}}\dd{\Phi_{n-1}(p-p_{n};p_{1},\cdots,p_{n-1})}.
\end{aligned}
\label{eq:phase-space-recursion}
\end{equation}
Let us define
\begin{equation}
k_{i}\equiv\sum_{j=1}^{i}p_{j}\thinspace,\ \ M_{i}^{2}\equiv k_{i}^{2}\thinspace,
\end{equation}
then we can insert the following integrals
\begin{align}
1 & \equiv\int\dd{M_{n-1}^{2}}\delta\qty(k_{n-1}^{2}-M_{n-1}^{2}),\\
1 & \equiv\int\dd[4]{k_{n-1}}\delta^{(4)}\qty(p-p_{n}-k_{n-1})
\end{align}
into Eq. (\ref{eq:phase-space-recursion}) and obtain 
\begin{equation}
\begin{aligned}\int\dd{\Phi_{n}} & =\int_{\mu_{n-1}^{2}}^{(M_{n}-m_{n})^{2}}\frac{\dd{M_{n-1}^{2}}}{2\pi}\\
 & \qquad\times\int\frac{\dd[4]{k_{n-1}}}{(2\pi)^{4}}\frac{\dd[4]{p_{n}}}{(2\pi)^{4}}(2\pi)^{4}\delta^{(4)}(p-p_{n}-k_{n-1})2\pi\delta\qty(k_{n-1}^{2}-M_{n-1}^{2})2\pi\delta\qty(p_{n}^{2}-m_{n}^{2})\\
 & \qquad\times\int\dd{\Phi_{n-1}(p-p_{n};p_{1},\cdots,p_{n-1})}\\
 & =\int_{\mu_{n-1}^{2}}^{\qty(M_{n}-m_{n})^{2}}\frac{\dd{M_{n-1}^{2}}}{2\pi}\int\dd{\Phi_{2}}(k_{n};k_{n-1},p_{n})\dd{\Phi_{n-1}(p-p_{n};p_{1},\cdots,p_{n-1})}\\
 & =\int_{\mu_{n-1}^{2}}^{\qty(M_{n}-m_{n})^{2}}\frac{\dd{M_{n-1}^{2}}}{2\pi}\int\frac{\dd{\Omega_{n-1}}}{(2\pi)^{2}}\frac{\sqrt{\lambda\qty(M_{n}^{2},M_{n-1}^{2},m_{n}^{2})}}{8M_{n}^{2}}\\
 & \qquad\times\int\dd{\Phi_{n-1}(p-p_{n};p_{1},\cdots,p_{n-1})}.
\end{aligned}
\end{equation}
where $\mu_{i}\equiv\sum_{j=1}^{i}m_{j}$, $m_{i}^{2}\equiv p_{i}^{2}$,
and $\dd{\Omega_{i}}$ is defined as the differential solid angle
in the rest frame of $k_{i+1}$, i.e., $\boldsymbol{k}_{i+1}=0$.

So the final result is
\begin{equation}
\int\dd{\Phi_{n}}=\frac{1}{(2\pi)^{3n-4}}\int_{\mu_{n-1}}^{M_{n}-m_{n}}\dd{M_{n-1}}\cdots\int_{\mu_{2}}^{M_{3}-m_{3}}\dd{M_{2}}\qty(\int\prod_{i=1}^{n-1}\dd{\Omega_{i}})\frac{1}{2^{n}M_{n}}\prod_{i=2}^{n}P_{i}\thinspace,\label{eq:n-body-phase-space-integration}
\end{equation}
where 
\begin{align}
P_{i} & \equiv\frac{\sqrt{\lambda\qty(M_{i}^{2},M_{i-1}^{2},m_{i}^{2})}}{2M_{i}}\thinspace,\\
\lambda(x,y,z) & \equiv(x-y-z)^{2}-4yz\thinspace.
\end{align}

Substituting Eq.~\eqref{eq:n-body-phase-space-integration} into Eq.~\eqref{eq:differential-decay-rate},
we can perform Monte Carlo integration straightforwardly, provided
that $|{\cal M}|^{2}$ is known. For those four- or five-body processes
considered in this work, $|{\cal M}|^{2}$ can be computed analytically
assuming the interaction in Eq.~\eqref{eq:-26}. The results for the
four-fermion operators are
\begin{align}
|{\cal M}_{S}|^{2} & =\frac{g_{{\rm eff}}^{2}}{M_{W}^{4}}\frac{4p_{4}.p_{5}}{\Lambda_{s}^{2}q^{4}}\left(M^{2}\left(p_{2}.p_{3}q^{2}-2q.p_{2}q.p_{3}\right)+2p.p_{2}\left(2p.qq.p_{3}-p.p_{3}q^{2}\right)\right),\label{eq:-27}\\
|{\cal M}_{V}|^{2} & =\frac{g_{{\rm eff}}^{2}}{M_{W}^{4}}\frac{16p_{3}.p_{4}}{\Lambda_{V}^{2}q^{4}}\left(M^{2}\left(p_{2}.p_{5}q^{2}-2q.p_{2}q.p_{5}\right)+2p.p_{2}\left(2p.qq.p_{5}-p.p_{5}q^{2}\right)\right),\label{eq:-28}\\
|{\cal M}_{T}|^{2} & =64\frac{g_{{\rm eff}}^{2}}{M_{W}^{4}}\frac{2p_{3}.p_{5}}{q^{4}}\left(M^{2}\left(p_{2}.p_{4}q^{2}-2q.p_{2}q.p_{4}\right)+p.p_{2}\left(4p.qq.p_{4}-2p.p_{4}q^{2}\right)\right)\\
 & +64\frac{g_{{\rm eff}}^{2}}{M_{W}^{4}}\frac{2p_{3}.p_{4}}{q^{4}}\left(M^{2}\left(p_{2}.p_{5}q^{2}-2q.p_{2}q.p_{5}\right)+p.p_{2}\left(4p.qq.p_{5}-2p.p_{5}q^{2}\right)\right)\\
 & +64\frac{g_{{\rm eff}}^{2}}{M_{W}^{4}}\frac{p_{4}.p_{5}}{q^{4}}\left(M^{2}\left(2q.p_{2}q.p_{3}-p_{2}.p_{3}q^{2}\right)+2p.p_{2}\left(p.p_{3}q^{2}-2p.qq.p_{3}\right)\right)\thinspace,
\end{align}
where the subscripts $S$, $V$, $T$ indicate the type of the four-fermion
operators used. For $S'$ and $V'$ operators, the results can be
simply obtained via ${\cal M}_{S'}={\cal M}_{S}|_{\Lambda_{s}\to\Lambda_{s'}}$,
${\cal M}_{V'}={\cal M}_{V}|_{\Lambda_{V}\to\Lambda_{V'}}$. 

For the scalar mediator in Eq.~\eqref{eq:-22}, the squared amplitude
of $^{8}\mathrm{B}\to\mathrm{^{8}Be}+e^{+}+\overline{\nu}+\phi$ is
given by
\begin{equation}
|{\cal M}_{\phi}|^{2}=\frac{g_{{\rm eff}}^{2}}{M_{W}^{4}}\frac{g_{\phi}^{2}}{q^{4}}\left(2M^{2}\left(p_{2}.p_{3}q^{2}-2q.p_{2}q.p_{3}\right)+p.p_{2}\left(8p.qq.p_{3}-4p.p_{3}q^{2}\right)\right),\label{eq:-29}
\end{equation}
where $q=p_{3}+p_{4}$ with $p_{3}$ and $p_{4}$ denoting the momenta
of $\overline{\nu}$ and $\phi$, respectively.  

\begin{acknowledgments}
We acknowledge the use of \textsc{FeynCalc}~\cite{Mertig:1990an,Shtabovenko:2016sxi,Shtabovenko:2020gxv}
and \textsc{PackageX}~\cite{Patel:2015tea,Patel:2016fam} in this
work.  X.-J.~Xu is supported in part by the National Natural Science
Foundation of China under grant No.~12141501 and also by the CAS
Project for Young Scientists in Basic Research (YSBR-099). Q.-f.~Wu
is supported by the National Natural Science Foundation of China under
grant No.~12075251.   
\end{acknowledgments}

\bibliographystyle{JHEP}
\bibliography{ref}

\end{document}